\documentclass[
]{ceurart}

\usepackage{listings}
\begin{document}

\copyrightyear{2023}
\copyrightclause{Copyright for this paper by its authors.
  Use permitted under Creative Commons License Attribution 4.0
  International (CC BY 4.0).}


\title{Key Factors Affecting European Reactions to AI in European Full and Flawed Democracies}


\author[1]{Long Pham}
\address[1]{Insight SFI Research Centre for Data Analytics, School of Computer Science, University College Cork, Ireland}

\author[1]{Barry O'Sullivan}
\address[2]{School of Computing, Dublin City University}

\author[2]{Tai Tan Mai}

\cortext[1]{Corresponding author: Tai Tan Mai - tai.tanmai@dcu.ie}
\fntext[1]{These authors contributed equally.}

\begin{abstract}
This study examines the key factors that affect European reactions to artificial intelligence (AI) in the context of both full and flawed democracies in Europe. Analysing a dataset of 4,006 respondents, categorised into full democracies and flawed democracies based on the Democracy Index developed by the Economist Intelligence Unit (EIU), this research identifies crucial factors that shape European attitudes toward AI in these two types of democracies. The analysis reveals noteworthy findings.
Firstly, it is observed that flawed democracies tend to exhibit higher levels of trust in government entities compared to their counterparts in full democracies. Additionally, individuals residing in flawed democracies demonstrate a more positive attitude toward AI when compared to respondents from full democracies. However, the study finds no significant difference in AI awareness between the two types of democracies, indicating a similar level of general knowledge about AI technologies among European citizens. Moreover, the study reveals that trust in AI measures, specifically ``Trust AI Solution,'' does not significantly vary between full and flawed democracies. This suggests that despite the differences in democratic quality, both types of democracies have similar levels of confidence in AI solutions.

\end{abstract}


\maketitle

\section{Introduction}
The rapid advancements of AI have revolutionised various sectors, including practices of democracy. While governments around the world increasingly embrace AI technologies to enhance their decision-making processes, improve service delivery, and address complex societal challenges, questions arise concerning the relationship between AI and democracy. Both AI and democracy encompass breadth and depth related to research domains of political science, public policies and administration, technology evolution and adoption, computer science, and many more.

The trajectories of AI and democracy, or democratic practices and institutions to be exact, are complex. Central to the successful adoption of AI in the public sector is the level of citizen trust in these technologies. Trust acts as a crucial foundation for democratic governance, and the application of AI has the potential to either bolster or erode citizen trust. Therefore, understanding the determinants of citizen trust in aspects of AI becomes essential in ensuring that the benefits of AI are realised without compromising democratic values. Globally, democracies are at different levels of key democratic features including electoral process and pluralism, functioning of government, political participation, political culture, and civil liberties~\cite{unit2022democracy}. Based on the key features, the Democracy Index,\footnote{\url{https://www.eiu.com/n/campaigns/democracy-index-2022/}} developed and sustained by the Economist Intelligence Unit, classifies the world's political systems into four regime types: full democracies, flawed democracies, hybrid regimes, and authoritarian regimes. Given the diversity in political systems in Europe, all four regimes exist. 

\section{Relevant Studies}
Democracy includes theoretical liberal, pluralistic, and deliberative models such as frameworks. However, at its core, democracy can be succinctly defined as the ``government of the people, by the people, for the people,'' as expressed in Lincoln's Gettysburg Address in 1863. According 
to~\cite{larry2008spirit}, democracy comprises four essential components: (1) a political system that facilitates the selection and replacement of the government through free and fair elections; (2) active and inclusive citizen participation in politics and civic life, including a vibrant public sphere and independent media; (3) safeguarding the human rights of all individuals; and (4) adherence to the rule of law, ensuring the separation of powers and equal application of laws and procedures to every 
citizen~\cite{larry2008spirit, schneider2020democratic}. In each of the four components, technologies have been applied at various paces and depths in all political regimes. One of the most recent technologies is digital, which was enabled by accessibility, affordability, availability, and others. Digital technologies enhance access to information, ensuring that citizens are well-informed. They enable citizens to express their views on projects and societal issues that affect them through consultations. Digital technologies engage citizens in decision-making processes, promoting democratic participation~\cite{oecd2001}. Many governments have implemented e-government services to optimize effectiveness and efficiency, prioritise citizen-centric design, and foster trust in governance~\cite{pena2020oecd}.
Within digital technologies, AI holds a prominent position. While introducing AI offers unparalleled opportunities to improve the efficiency and effectiveness of public action, governments must also ensure that it aligns with the core values of liberal democracies~\cite{sharma2020artificial}. 

Although AI and its applications can be found in many different domains such as smart cities~\cite{pham2016key}, education~\cite{mai2019students} and 
FinTech~\cite{nguyen2022cryptocurrency}, studies on citizen perceptions of AI have been scarce~\cite{konig2022citizen}, especially those in empirical research. In a study that utilises original data from a representative German online panel and aims to understand whether different democratic conceptions influence citizen acceptance of 
AI~\cite{konig2022citizen}, detected some interesting findings. The study found citizens who support a liberal-democratic conception of democracy will be less supportive of the use of AI in government and politics. Citizens who support a technocratic conception of democracy will be more supportive of the use of AI.
The study highlights that overall evaluations of AI and support for its use in politics are relatively positive across Europe, including Germany. Citizens tend to be sceptical of using AI in high-level political decision-making and their conceptions of democracy are found to be relevant predictors of AI support. The study also finds that women are less supportive of AI in public administration decision-making, while individuals with higher formal education tend to be more supportive. 
Apart from these relevant studies, citizen trust in governmental entities and political systems, their awareness, and their attitude toward technology usage in public services are a mixed bag with more positive outcomes from e-government applications~\cite{pena2020oecd, duberry2022artificial}. Deriving from the existing findings in democracies, perceptions of AI, citizen awareness and attitudes toward AI, this paper explores the following hypotheses:
\begin{itemize}
    \item \textit{H1.}~There are differences between full/flawed democracies in trust variables (e.g., trust in governmental/authority entities).
    \item \textit{H2.}~There are differences between full/flawed democracies in overall attitude toward AI. 
    \item  \textit{H3.}~There are differences between full/flawed democracies in overall awareness of AI.
    \item \textit{H4.}~There are differences in the relationships of the key demographic variables of age, gender, and education with relevant variables of awareness, attitude, and trust in AI between full and flawed democracies.
\end{itemize}
\section{Research Method}
\subsection{Questionnaire and Data Collection}
The questionnaire was developed by \cite{scantamburlo2023artificial}. It comprised 14 items covering the dimensions of trust, awareness, and attitude. The questionnaire utilised a Likert scale, with a range of 1 to 5 where 1 - negative/low values and 5 - positive/high values. For a thorough description of the questionnaire, the full text and the data collection process can be seen in \cite{scantamburlo2023artificial}.
In the scope of this study, we merely analyse specific Likert-scale questions related to AI Awareness (Q7), AI Attitude (Q8), Trust in AI Solutions (Q12), and Trust in Entities (Q14). Demographic factors such as age, gender and education were also included. The survey was conducted via online interviews in June 2021, with an average completion time of 20 minutes. The majority of the respondents live continuously in their countries. Respondent data was anonymised and handled in accordance with the European Union's General  Data Protection Regulation (GDPR). 

Additionally, respondents were labelled as coming from either a \textit{full} or \textit{flawed democracy} regime based on the country where they were interviewed. The democracy classification of each country was extracted from~\cite{unit2022democracy}. 
  
\subsection{Statistical Analysis}

A statistical analysis to assess the validity and reliability of the questionnaire was performed similarly in \cite{scantamburlo2023artificial}. Our analysis involved conducting both Exploratory Factor Analysis (EFA) and Confirmatory Factor Analysis (CFA) to evaluate the reliability of the questionnaire items. The sample (n=4,006) was randomly equally divided into two groups: n1=2,003 for EFA and n2=2,003 for CFA. Prior to conducting the EFA, we evaluated the sample adequacy using the Kaiser-Meyer-Olkin (KMO) test  and Bartlett's test of sphericity \cite{kaiser1974index}.  To evaluate the internal consistency of the EFA solution, we used Cronbach's $\alpha$ \cite{gadermann2019estimating}. The CFA was conducted using a polychoric matrix and the diagonally weighted least squares (DWLS) extraction method \cite{li2016confirmatory}. 

In addition, Bivariate analysis was then employed to examine potential significant relationships between identified factors and the demography variables. Non-parametric tests, specifically the Mann-Whitney U (where two groups) \cite{mann1947test} and Kruskal-Wallis H (where more than two groups) \cite{kruskal1952use} tests, were conducted considering its non-normal distribution. 

\section{Results and Discussion}

The survey data consisted of responses from 4,006 individuals across 8 European countries. According to \cite{unit2022democracy}, respondents in Italy, Romania and Poland were labelled as  \textit{flawed democracy} while those in the Netherlands, Sweden, Spain, Germany and France were classified as \textit{full democracy}. The respondents also represented diverse demographics, including 1,976 females, 2,019 males, 1,107 young (18-34 years old), 1,578 middle-aged (35-54 years old), and 1,321 older adults (55-75 years old). The questionnaire data summary can be found in \cite{dataSummary}.

\subsection{Questionnaire data validation and Factor Analysis}
The data used for EFA showed high suitability, as indicated by a KMO = 0.96 > 0.7 and a significant p-value from Bartlett's test of sphericity, aligning with previous research \cite{kaiser1974index}.
Using parallel analysis \cite{lim2019determining}, we identified four factors, namely \textit{AI Awareness}, \textit{AI Attitude}, \textit{Trust in Governmental Entities}, and \textit{Trust in AI Solutions}, that accounted for 62\% of the total variance. Positive correlations were found among all four factors, while none exceeded the threshold of 0.7, indicating satisfactory discriminant validity.
Cronbach's alpha values exceed 0.8, indicating good internal reliability and consistency. Additionally, CFA was conducted to evaluate the proposed factorial structure. The CFA results supported the reliability and validity of the instrument, with acceptable fit indices (RMSEA = 0.011, CFI = 0.995, TLI = 0.994, SRMR = 0.03, p < 0.0001) \cite{kline2023principles}. Then, the identified factors were used in bivariate analysis to validate the hypotheses reported and discussed below.

   \begin{table}
    \begin{center}
    {\caption{Bivariate analysis results between democracy regime and identified factors. $\Box$ - Mean; $\bigtriangleup$ - Median}\label{fullData}}
    \begin{tabular}{llllll}
    \cline{1-6}
    \multicolumn{1}{c}{} & \multicolumn{1}{c}{\textbf{N}} & \multicolumn{1}{p{2.5cm}}{\textbf{Trust in Gov Entities}}         & \multicolumn{1}{p{2.5cm}}{\textbf{Trust in AI Solutions}} & \multicolumn{1}{p{2.5cm}}{\textbf{AI Awareness}} & \multicolumn{1}{p{2.5cm}}{\textbf{AI Attitude}}                  \\ \cline{1-6}
    
    \textbf{Mann–Whitney U}                              &                                & \multicolumn{1}{c}{\textbf{p-value \textless 0.0001}}             & \multicolumn{1}{c}{\textbf{p-value = 0.0002}}  & \multicolumn{1}{c}{p-value = 0.117}  & \multicolumn{1}{c}{\textbf{p-value \textless 0.0001}}            \\
    \cline{1-6}
    Flawed democracy                              & \multicolumn{1}{c}{1501}       & \multicolumn{1}{c}{$\Box$: 3.82, $\bigtriangleup$: 3.89}                        & \multicolumn{1}{c}{$\Box$: 4.02, $\bigtriangleup$: 4.00}    & \multicolumn{1}{c}{$\Box$: 3.65, $\bigtriangleup$: 3.70}  & \multicolumn{1}{c}{$\Box$: 3.85, $\bigtriangleup$: 4.00}                      \\
    Full democracy                                         & \multicolumn{1}{c}{2505}       & \multicolumn{1}{c}{$\Box$: 3.64, $\bigtriangleup$: 3.70}                        & \multicolumn{1}{c}{$\Box$: 3.91, $\bigtriangleup$: 4.00}      
    & \multicolumn{1}{c}{$\Box$: 3.57, $\bigtriangleup$: 3.70}
    & \multicolumn{1}{c}{$\Box$: 3.54, $\bigtriangleup$: 3.60}
    \\
    \cline{1-6}
       \multicolumn{2}{l}{Hypothesis support}    &  \textbf{H1 supported} & \textbf{H1 supported} & H3 not supported & \textbf{H2 supported}

    \end{tabular}
    \end{center}
    \vspace{-0.5cm}
    \end{table}

 \begin{table}
    \begin{center}
    {\caption{Bivariate analysis results between demographic and identified factors in respondents from flawed democracy and full democracy regimes. * = p-value < .05, ** = p-value < .001, *** = p-value <  .0001, empty value = not significant (p-value >  .05).}\label{splitdata}}
    \begin{tabular}{llllll}
    \cline{1-6}
    \multicolumn{1}{c}{} & \multicolumn{1}{c}{\textbf{}} & \multicolumn{1}{p{2.5cm}}{\textbf{Trust in Gov Entities}}         & \multicolumn{1}{p{2.5cm}}{\textbf{Trust in AI Solutions}} & \multicolumn{1}{p{2.5cm}}{\textbf{AI Awareness}} & \multicolumn{1}{p{2.5cm}}{\textbf{AI Attitude}}                  \\ 
    \cline{1-6}
    \multicolumn{6}{c}{\textbf{Flawed democracy regime respondents}}                                                                    \\
         \cline{1-6}
     \textbf{Education level}                     &                                & \multicolumn{1}{c}{\textbf{*}} & \multicolumn{1}{c}{\textbf{***}} &    &  \textbf{***}  \\
    \cline{1-6}

    \textbf{Gender}                              &                                & \multicolumn{1}{c}{\textbf{*}}             & \multicolumn{1}{c}{\textbf{*}} & \multicolumn{1}{c}{\textbf{*}}   &          \\
    \cline{1-6}

    \textbf{Age group}                           &                                & \multicolumn{1}{c}{}           & \multicolumn{1}{c}{\textbf{***}} &    &   \\
    \cline{1-6}\\

        \multicolumn{6}{c}{\textbf{Full democracy regime respondents}}                                                                                 \\
        \cline{1-6}
     \textbf{Education level}                     &                                & \multicolumn{1}{c}{\textbf{***}} & \multicolumn{1}{c}{\textbf{***}} & \multicolumn{1}{c}{\textbf{***}} & \multicolumn{1}{c}{\textbf{**}} \\
    \cline{1-6}

    \textbf{Gender}                              &                                & \multicolumn{1}{c}{}             & \multicolumn{1}{c}{}  &   & \textbf{**}         \\
    \cline{1-6}

    \textbf{Age group}                           &                                & \multicolumn{1}{c}{}           & \multicolumn{1}{c}{\textbf{***}} &   & \textbf{**} \\
    \cline{1-6}
       \multicolumn{2}{l}{Hypothesis support}    &  \multicolumn{4}{c}{\textbf{H4 partly supported}}
                                                                 
    \end{tabular}
    \end{center}
    \vspace{-1cm}
    \end{table}

\subsection{Analysis \& Discussion}
Table \ref{fullData} and \ref{splitdata} present detailed bivariate analyses based on the collected data sets for the specific aims of this study, where discover insights on the full/flawed democracies of eight EU countries.

\textit{Hypothesis 1} was confirmed (Table~\ref{fullData}). Responses from flawed democracies (e.g., Italy, Romania, and Poland) showed that people there tend to have higher levels of trust in both governmental/authority entities and trust in measures ensuring safe usage and better-regulated AI applications. Meanwhile, those from full democracies (e.g., France, Germany, Sweden, Netherlands, and Spain) tend not to have a positive response. These findings verified the notions suggested that citizens who support a liberal-democratic conception of democracy were less supportive of the use of AI in government and politics \cite{konig2022citizen} because they have less trust in and are more sceptical of governments. Trust in government among democratic systems has faced significant challenges \cite{geana2011health}. There has been a decrease in public trust not only in the government but also in business, public media, and non-governmental organisations (NGOs)~\cite{geana2011health}. Consequently, this decline might foster a sense of unfairness and powerlessness and a diminished belief and trust in the current system. However, the response to the global pandemic had a significant impact on the level of trust people had in their governments. A study conducted 
by~\cite{groeniger2021dutch} on trust in government during the pandemic revealed that when there is a higher level of trust in the government, people are more likely to comply with health-related policies. The trust that the public places in their government and political authorities plays a crucial role in effectively implementing policies related to public health and safety.

\textit{Hypothesis 2} was supported 
(Table~\ref{fullData}). Those from flawed democracies tend to have a higher positive attitude toward AI than those from full democracies. The findings are confounding under the levels of trust demonstrated by full democracies, more sceptical and less trusting, but contradictory under the correlations between the overall attitude of flawed democracies, which were not performing that well in the electoral process and pluralism, functioning of government, political participation, political culture, and civil liberties.

\textit{Hypothesis 3} was not supported (Table \ref{fullData}). This was unexpected because full democracies tend to do very well in the five key categories of the electoral process and pluralism, functioning of government, political participation, political culture, and civil liberties. One of the key factors for doing well in those is the high level of awareness among citizens about policies, practices, services, and political activities by governments, authorities, and parties, thus citizens
can participate in democratic practices and institutions.

\textit{Hypothesis 4} was partially supported (Table \ref{splitdata}). In fact, education has positive correlations with trust in entities and AI measures and with positive attitudes toward AI in both full and flawed democracies. The only difference in education between the two groups was no correlation in awareness among the flawed democracies. In the age variable, the two groups shared a positive correlation with trust in measures to ensure the safe use of AI while the full democracies also have a positive correlation with a supportive attitude toward AI.

In the gender variable, the two genders in flawed democracies had no difference in awareness, trust of entities, and AI measures. The only difference between the two genders in this group of countries is their attitude toward AI. On the contrary, in the full democracies, there was only a difference between the two genders in AI attitude, which means females in these countries tend to have a positive attitude toward AI while there was no difference between the two genders in awareness of AI and the two trust variables. These findings conflicted with the results of the study by \cite{konig2022citizen} showed that women are less supportive of AI in public administration decision-making, while individuals with higher formal education tend to be more supportive.

The connection between trust and citizen attitudes toward AI and political institutional design is crucial when it comes to the implementation of AI technology in a manner that empowers and enhances trust for citizens. Transparency, accountability, citizen participation, ethical design, bias mitigation, and education are all important dimensions that political institutions should consider when shaping AI policies and practices.

\section{Conclusions}
The complex relationships between AI, democracy, citizen awareness, attitude, and trust in democratic practices were partly demonstrated by the analyses of the full democracies and flawed democracies in Europe. Increasingly, AI applications have become integrated into democratic institutions, processes, and activities including electoral process and pluralism, functioning of government, political participation, political culture, and civil liberties. This study is one of the first attempts at bringing empirical pieces of evidence to light for many relevant aspects of the association between citizens, AI, and democracy in at least two democratic systems of full democracies and flawed democracies. The goal is to find the key factors and appropriate pathways to make a positive influence on European reactions to AI and enhance those indicators that uplift democratic practices and institutions of flawed democracies to full democracies or elevate even higher standards in those in the already full democracies. 
Embracing some limitations, this study contributes to the growing literature with insights and findings emphasising the importance of democratic contexts when analysing the implications of AI on citizen perceptions and suggesting the need for well-informed and inclusive AI governance in democratic societies.

\section{Acknowledgements}
This research has received funding from the European Union’s Horizon 2020 research and innovation programme under grant agreement No. 825619 and a grant from Science Foundation Ireland under Grant number 12/RC/2289-P2 at Insight the SFI Research Centre for Data Analytics at UCC, which is co-funded under the European Regional Development Fund. The data collection for this research was supported by Ca' Foscari University. This work reflects only the authors' view and the Commission is not responsible for any use that may be made of the information it contains.

\bibliography{sample-ceur}

\appendix

\end{document}